\documentclass{ws-procs975x65}

\begin{document}

\title{VENUS TRANSITS: HISTORY AND OPPORTUNITIES FOR PLANETARY, SOLAR AND GRAVITATIONAL PHYSICS}

\author{C. SIGISMONDI,$^1,^2$ X. WANG,$^3$  P. ROCHER,$^4$  E. REIS-NETO$^2$}

\address{$^1$ICRANET,Sapienza University and 
Pontifical Atheneum Regina Apostolorum, Rome, Italy sigismondi@icra.it 
$^2$Observatorio Nacional, Rio de Janeiro, Brazil  eugenio@on.br
$^3$National Astronomical Observatories of China and Chinese Academy of Science, Huariou Solar Observing Station, China wxf@naoc.cn
$^4$IMCCE Paris, France rocher@imcce.fr}

\begin{abstract}
The data of 2012 transit of Venus are compared with the ones of 2004.
The thickness of the atmosphere of Venus, its aureole and the effect of oblateness and other asphericities in the figure of the Sun are taken into consideration, 
as well as the black drop effect. A new extrapolation method for the contact times is presented.
The next Mercury transit in 2016 will be fully visible from Europe, and the data will be gathered in view of this new method of analysis, to obtain the solar diameter. 
\end{abstract}

\keywords{Venus Transit, Venus atmosphere, Black-drop effect, Solar diameter.}

\bodymatter

\section{Studies on the distance and the figure of the Sun}
The first opportunity offered by the transit of Venus was the measurement of the Astronomical Unit, already perfectly known in 1874 and 1882 transits 
as reading the reports of S. Newcomb:\cite{newcomb} the solar parallax was measured from the contact timings of the transits of Venus as $8.794\pm0.018$", 
being $8.794143$" the current value adopted in the astronomical constants. The Astronomical Unit is the fundamental distance for all astronomy and astrophysics 
and for gravitational physics to determine the product $G \times M_\odot$ using the Newton and Kepler's laws.

The measurement of variations the solar diameter and of the figure of the Sun has important astrophysical and relativistic implications.\cite{sigi11} 
In the first half of the 20th century, the figure of the Sun was studied in relation to solar activity. 
With the detection of a deficit of the neutrino flux from the Sun during the Brookhaven $^{37}$Cl experiment, 
with the development of gravitational theories alternative to general theory of relativity, and with the progress made
in helioseismology, astrometric studies of the Sun have intensified sharply. Of particular interest are the
possible secular and long-period components in the oscillations of the solar radius and its variations with a period close to the 
11-year solar cycle.\cite{svesh} 

\section{Data from transits of 2004 and 2012 in comparison}
In 2004 the attention of the scientific world to the transit of Venus was concentrated to the study of the aureole, with very good observations
from Themis, made by the late Jean Arnaud (1942-2010). 
The project VenusTEX in 2012 coordinated by Paolo Tanga has taken this heritage: nine observing stations spread worldwide and equipped 
with special off-axis coronographs designed to cut off the light of photosphere in order to see the aureole of Venus cast 
over the inner corona and immersed in the halo around the Sun,
as seen from ground, during the ingress and the egress of Venus over the solar disk.\cite{tanga}
The transit of 2012 was the last chance to see a planet with atmosphere across the Sun before the year 2117, and this phenomenon is being carefully studied also 
to obtain good models to interprete the forthcoming observations of extrasolar planets transits with similar properties.

\section{Solar Astrometry with planetary transits}
Mercury has a smaller parallax than Venus, and did not help for the measurement of the Astronomical Unit as Venus,
but their transits are as frequent as 11-12 per century, 
and they have been used in 1980 to monitor the diameter of the Sun, starting from 1736.\cite{shapiro}
The analysis has been extended in 2002 back to 1631 with 4500 contact timings analyzed.\cite{svesh}
This series of data is affected by the black-drop effect which is primarily related to the diameter of the telescope used in the observations.\cite{sigi08} 
Now the last two Venus transits offer the possibility to compare new data 
with modern observational standards to overcome the problems due the black-drop effect.
For this purpose are necessary chronodated images of the transit, containing both ingress and egress. 
This requirement excludes the majority of the images available from both transits of 2004 and 2012.
After a wide research only 50 images have been found for 2004, 25 at ingress and 25 at egress taken by Anthony Ayomamitis in Athens with a 6 cm optical system
composed by Coronado H$_{\alpha}$-line filter and an apochromatic refracting telescope. Despite of a strong black-drop effect a new method of data reduction promises to overcome it almost completely. 
This equipment was equivalent to the one used by Joseph Gambart during the Mercury transit of 1832,\cite{gambart} 
arising problems already discussed,\cite{sigi08} but an extrapolation of the contact timings is now possible by inspecting the images recorded.
For 2012 at the Huairou Solar Observing Station we organized the observation of the transit getting one image per second with 12-bits dynamics, i.e. 4096 levels of intensity, 4 mega-pixel and covering all phases of ingress and egress.
About the satellites, in 2012 HINODE, SDO, RHESSI and PICARD made images of the transit while SOHO, which is in the first Lagrangian point of Earth's orbit L1, both Venus transits were unvisibile, while it observed Mercury transits in 2003 and 2006.\cite{emilio}

\section{Black-drop effect}

The black drop effect is an effect mainly dependent on the instrumental point spread function which plays the role of 
the penumbra in the similar phenomenon of the attraction of the shadows.\cite{lang}
Second order contributions come from the astigmatisms of the instrument 
[{\rm http://www.bo.astro.it/\~biblio/Horn/dicembre3.htm}]\cite{horn} 
and from the convolution with the solar limb darkening.\cite{pasa}
On small aperture telescopes $\emptyset \le 10$cm this effect is large, but it tends to disappear in larger telescopes 
$\emptyset \ge 25$cm.
It affects the observations of timing contacts of planetary transits and also of Baily's beads 
merging during total eclipses.
The determination of the contacts between Venus and solar limb can be better obtained as an extrapolation to zero of
the intersection between Venus and solar disks. This is made by fitting the analytic formula which calculates these
intersections to their measured lengths, during ingress and egress phases.
Other usefull reference times are the two times when these intersections are maxima, respectively at ingress and egress.
The comparison between ephemerides and observations-extrapolations of these times gives the variation of the solar radius,
through the proportion
$\Delta T_{calc} \backslash \Delta T_{obs}= 959.63" \backslash R_{\odot}$,
where 959.63" is the standard solar radius of 696000 Km at 1 AU, corresponding to the one measured by Auwers in 1891.\cite{auwers}

\section{Solar figure corrections}
The figure of the Sun is not perfectly spherical. Its oblateness has been the object of several observational campaigns in order to assess the relativistic contribution to the perihelion precession of Mercury, and distinguish it from the classical quadrupole momentum $J_2$.\cite{sigi05}
One of the latest results in this field has been obtained by RHESSI satellite\cite{rhessi} and it shows the Sun with a small oblateness of 6 km, and minor departures from this ellipsoid. This value is compatible with the solar rotational rate, while the oblateness has been recently claimed as stable during the solar cycle.\cite{emilio2} Further investigations on this topic are deserved because global oscillations of the figure of the Sun provide additional information on its internal structure.
 
\section{Ephermerides: trustability and accuracy}

The ephemerides INPOP06\cite{fienga} used in the computation of $\Delta T_{calc}$ are developped at IMCCE with accuracy as good as 0.001 s in timing, for the centers of mass of the involved bodies. The reference radius of Venus and Sun are the ones in the astronomical constants. Observational accuracy in timing of 0.1 s would allow a resolution in 
$\Delta R_{\odot}$ of one part over 200000, i.e. 0.01".





\begin{thebibliography}{9}

\bibitem{newcomb} S. Newcomb {\em The elements of the four inner planets and the fundamental constant of astronomy}, 157 (Washington DC, 1895).
\bibitem{sigi11} C. Sigismondi, {\em Int. J. Mod. Phys. Conf. S.}, {\bf 3}, 464 (2011).
\bibitem{svesh} M. L. Sveshnikov, {\em Astronomy Letters}, {\bf 28}, 115 (2002).
\bibitem{tanga} P. Tanga, et al., {\em ICARUS}, {\bf 218}, 207 (2012).
\bibitem{shapiro} I. I. Shapiro, {\em Science}, {\bf 208}, 51 (1980).
\bibitem{sigi08} C. Sigismondi, {\em AIP Conf. Proc.}, {\bf 1059}, 189 (2008).
\bibitem{gambart} J. Gambart, {\em Astron. Nach.}, {\bf 10}, 257 (1832).
\bibitem{emilio} M. Emilio, et al. {\em Astrophys. J.}, {\bf 750}, 135 (2012).
\bibitem{lang} F. Lang da Silveira and R. Axt {\em F\'isica na Escola}, {\bf 8}, 17 (2007).
\bibitem{horn} G. Horn d'Arturo, {\em Pub. Oss. Astron. R. Univ. Bologna}, {\bf 1}, (1922).
\bibitem{pasa} J. Pasachoff, G. Schneider and L. Golub, {\em Proc IAUC}, {\bf 196}, 242 (2004).
\bibitem{auwers} A. Auwers, {\em Astron. Nach.}, {\bf 128}, 361 (1891).
\bibitem{sigi05} C. Sigismondi, {\em Nuovo Cimento B}, {\bf 120}, 1169 (2005).
\bibitem{rhessi} M. D. Fivian {\it et al.}, {\it Science} {\bf 322}, 560 (2008). 
\bibitem{emilio} M. Emilio, et al. {\em Science}, {\bf 337}, 1638 (2012).
\bibitem{fienga} A. Fienga et al., Astron. and Astropys. {\bf 477}, 315 (2008).


\end{thebibliography}

\end{document}